\documentstyle[11pt,aaspp4]{article}
\journalid{}{}
\articleid{}{}
\def\lesssim{\mathrel{\hbox{\rlap{\hbox{\lower4pt\hbox{$\sim$}}}\hbox{$<$}}}}
\def\gtrsim{\mathrel{\hbox{\rlap{\hbox{\lower4pt\hbox{$\sim$}}}\hbox{$>$}}}}
\def\sq{\hbox{\rlap{$\sqcap$}$\sqcup$}}
\newcommand{\ropt}{R$_{25}$ }
\newcommand{\ha}{H$\alpha$ }
\begin{document}
\title{DISCOVERY OF RECENT STAR FORMATION IN THE EXTREME OUTER REGIONS OF
DISK GALAXIES}
\author{Annette~M.~N. Ferguson\altaffilmark{1,2,3} and Rosemary 
F.G. Wyse\altaffilmark{2,3}}
\affil{Department of Physics and Astronomy, The Johns Hopkins University,
    Baltimore, MD 21218}
\author{J. S. Gallagher}
\affil{Department of Astronomy, University of Wisconsin, 
Madison, WI 53706}
\author{Deidre A. Hunter}
\affil{Lowell Observatory, 1400 W. Mars Hill Road, Flagstaff, AZ 86001}
\altaffiltext{1}{ Current Address: Institute of Astronomy, University of
Cambridge, Madingley Road, Cambridge, UK CB3 0HA}
\altaffiltext{2}{Visiting Astronomer, Kitt Peak National Observatory. 
KPNO is operated by AURA, Inc.\ under contract to the National Science
Foundation.}
\altaffiltext{3}{Visiting Astronomer, Lowell Observatory.}

\begin{abstract}
We present deep H$\alpha$ images of three nearby late-type spiral
galaxies (NGC~628, NGC~1058 and NGC~6946), which reveal the presence of
HII regions out to, and beyond, two optical radii (defined by the 25th
B-band isophote).  The outermost HII regions appear small, faint and
isolated, compared to their inner disk counterparts, and are
distributed in organized spiral arm structures, likely associated with
underlying HI arms and faint stellar arms.   The relationship between
the azimuthally--averaged H$\alpha$ surface brightness (proportional to
star formation rate per unit area) and the total gas surface density is
observed to steepen considerably at low gas surface densities.   We
find that this effect is largely driven by a sharp decrease in the
covering factor of star formation at large radii, and not by changes in
the rate at which stars form locally.   An azimuthally--averaged
analysis of the gravitational stability of the disk of NGC~6946 reveals
that while the existence of star formation in the extreme outer disk is
consistent with the  Toomre-$Q$ instability model, the low rates observed
are only compatible with the model when a constant gaseous
velocity dispersion is assumed.  We suggest that observed behaviour
could also be explained  by a model in which the star formation rate
has an intrinsic dependence on the azimuthally-averaged gas {\it
volume\/} density, which decreases rapidly in the outer disk due to the
vertical flaring of the gas layer.

\end{abstract}


\section{Introduction}

An understanding of the processes which drive large-scale  star
formation in galactic disks is of great importance for models of disk
formation and evolution.  The  present-day star formation rates in the
previously poorly explored, faint outer regions of disks (i.e.  R
$\gtrsim$~\ropt, where \ropt is defined by the 25th magnitude B-band
isophote) provide important and unique insight on the nature of the
star formation law in disk galaxies.  Competing models make definite
and distinct predictions for the expected behavior in these parts (eg.
Prantzos \& Aubert 1995, Tosi 1996).  Characterized by their  low gas
surface densities (yet high gas fractions), low metallicities and long
dynamical timescales, these outer regions provide a probe of star
formation in physical environments which are very different from those
typical of the bright, inner regions of disks, but are  similar to
those inferred for many high-redshift  damped Lyman-$\alpha$ systems,
(eg. Pettini et al 1997), as well as giant low surface brightness
galaxies (eg. Pickering et al 1997).

In order to investigate the past and present star formation rates in
the extreme outer regions of disks, we have obtained wide field-of-view
(FOV), deep H$\alpha$ and broadband images for  a sample of $\sim$20
nearby disk galaxies.  In this Letter, we present first H$\alpha$ 
results for 3
galaxies (NGC~628 (D$=$10.7~Mpc), NGC~1058 (D$=$10.0~Mpc), NGC~6946
(D$=$5.3~Mpc)) where we have discovered HII regions, hence recent
massive star formation, at particularly large radii.  These late-type
spirals are all characterized by larger than average HI-to-optical
sizes.  The results and interpretation from the full sample  will be
presented in Ferguson et al (1998a,b).  Chemical abundance measurements
for several of our newly-discovered outer-disk HII regions  are
presented in Ferguson et al (1998c).

\section{Observations and Data Analysis}

Wide-field CCD observations were obtained for our  galaxies using the
KPNO 0.9~m (NGC~628, NGC~6946) and the Lowell 1.8~m (NGC~1058).  At
KPNO. we used the Tek 2048x2048 CCD at the f/7.5 Cassegrain  focus,
with a FOV of 23{\arcmin} x 23{\arcmin}, and 0.69\arcsec/pixel).  At
Lowell, we used the TI 800x800 CCD with the Ohio-State Imaging
Fabry-Perot Spectrometer  as  5:1 focal reducer, providing a FOV $\sim$
7{\arcmin} on a side, with 0.49\arcsec/pixel). Image reduction was
carried out in the standard manner (see Ferguson et al 1996).  The
radial variation of azimuthally--averaged H$\alpha$ surface brightness
was determined using elliptical aperture photometry on the H$\alpha$
images.   It was possible to do surface photometry on our images only
out to approximately R$_{25}$, at which point errors in the background
dominated the signal and often conspired to produce negative total
fluxes in a given annulus despite the fact that HII regions (albeit
faint)  would be clearly visible on our images.  To circumvent this
problem, we counted emission only above a local isophotal threshold.
This approach was equivalent to carrying out areal surface photometry
in the inner disk, where the covering fraction of emission above our
threshold surface brightness was high, and to carrying out  photometry
of discrete HII regions in the far outer disk.  The typical threshold
employed, always several times larger than the large-scale flat--field
error, was $\sim$
3~$\times$~10$^{-17}$~erg~s$^{-1}$~cm$^{-2}$~/$\Box^{\arcsec}$, which
is considerably lower than those usually used to define the boundaries
of discrete HII regions (e.g.  Kennicutt 1988; Feinstein 1997) and also
lower than the typical surface brightnesses exhibited by
bright-moderate diffuse ionized gas  (eg.  Ferguson et al 1996).  This
method of counting pixels only where there is clearly emission allows
us to get around the limitations imposed by the  flat--field errors,
but it should be kept in mind that it provides {\it strictly a lower
limit} to the mean \ha surface brightness in each annulus.  A full
discussion of the analysis will be included in Ferguson et al (1998a).

H$\alpha$ continuum--subtracted images for the galaxies are
shown in Figure 1.

\section{Discovery of Recent Star Formation in the Extreme Outer Limits
of Disks}

An exciting  and unexpected result is the discovery of recent massive
star formation in the extreme outer limits of  all three galaxies.
While early photographic studies of two of them revealed the existence
of HII regions out to slightly beyond the optical radius (eg. Hodge
1969, Bonnarel et al 1986), our data indicate the presence of HII
regions out to the extent of our imagery, i.e.  to more than two
optical radii.

Our images show a striking difference between the inner and outer disk
HII regions, in that  star formation the outer disk occurs in  smaller,
fainter and more isolated HII regions. The brightest outer disk HII
regions detected here have diameters of 150--500 pc and \ha
luminosities  of only 1--80~$\times$~10$^{37}$~erg~s$^{-1}$.  Assuming
that the HII regions are ionization-bounded (which may well not be the
case, e.g. Ferguson et al 1996) and  that they have negligible internal
extinction (see Ferguson et al 1998c), the stellar Lyman continuum
fluxes from Vacca et al (1996) predict a typical (lower limit to the)
ionizing population of 0.2--20 equivalent O5V stars.  Establishing
whether the populations of inner and outer disk HII regions are indeed
intrinsically different, or this merely reflects poor statistical
sampling  of the HII region luminosity function in the outer disk, will
be a difficult, yet important, task.

The degree of organization of the outer disk star formation is also
remarkable.  The outer HII regions are not randomly distributed, but
rather are in long, narrow spiral arms,  more obviously in NGC~628 and
NGC~6946. Our deep broad-band images reveal the existence of faint
(B~$\sim$~26--28 mag/$\sq$\arcsec) stellar arms in all three galaxies,
associated with these HII arms (Ferguson et al 1998a).  Furthermore,
inspection of published HI maps also reveals outer spiral arms in the
underlying neutral gas (eg. Shostak \& van der Kruit 1984; Dickey et al
1990; Kamphuis 1993).  The relationship between the inner and outer
spiral structure remains unclear, as indeed is the dynamics underlying
the outer arms. The lack of obvious companions to any of these galaxies
makes the tidal hypothesis for spiral arm formation unlikely; future
observations in the near-IR will help to distinguish between
alternative theories for the outer spiral patterns, such as long--lived
density waves, or transient shearing perturbations.

The existence of massive stars in the extreme outer regions of disks
means that stellar feedback is also active in these parts. The typical
outer disk star formation rates are $\sim$
0.01--0.05~M$_{\odot}$~pc$^{-2}$~Gyr$^{-1}$, giving Type II
supernova rates of $\sim$~1--4$\times$10$^{-4}$~SN~pc$^{-2}$~Gyr$^{-1}$
(assuming a Salpeter IMF, with
0.1~M$_{\odot}\lesssim$M$\lesssim$100~M$_{\odot}$, and that all stars
with M$\gtrsim$8~M$_{\odot}$ explode as supernovae).   Reasonable
estimates of their coupling efficiency to 
the ISM suggest that massive stars are likely to play a significant role
in the energy balance of the outer disk (but not in supporting the HI 
velocity dispersion; Sellwood and Balbus 1998).

\section{Discussion}

\subsection{The Rate of Star Formation in Outer Galactic Disks}

The radial behaviour of the azimuthally--averaged \ha surface
brightness distribution ($\Sigma_{H\alpha}$) traces the  mean
massive-star formation rate per unit area across the disk.  Figure 2
(left panel) presents the deprojected (corrected for cos{\it i}),
azimuthally--averaged radial  profiles of $\Sigma_{H\alpha}$, as
derived above.   These profiles are corrected for [NII] contamination
and Galactic extinction, but not for extinction within the galaxies
themselves.  Figure 2 (right panel) shows the relationship between the
azimuthally--averaged total gas surface density  (atomic plus molecular
gas with corrections for heavy elements) and the \ha profiles.

The azimuthally--averaged \ha surface brightness declines with
increasing radius, but with evidence for a change in the behavior of
the gradient in the outer disk, in all galaxies.  The surface
brightness profiles of NGC~628 and NGC~6946 both exhibit a sharp
fall-off in $\Sigma_{H\alpha}$ near the edge of the optical disk, with
a subsequent flattening of the decline at larger radii.   Steepenings
are also seen at low gas column densities in plots of
$\Sigma_{H\alpha}$--$\Sigma_{gas}$, where they occur at
azimuthally-averaged total gas surface densities of
5--10~M$_{\sun}$/pc$^2$,  close to that estimated to be required to
shield the molecular cloud cores from photodissociation by the ambient
UV radiation field (eg.  Federman et al 1979).  Indeed the location of
the steepenings is approximately coincident to the radius where the
disk undergoes the transition from being dominated by (warm) molecular
to atomic gas.

The existence of abrupt declines in the azimuthally--averaged H$\alpha$
surface brightnesses, hence star formation rates, at low gas surface
densities make it impossible to describe the observations with a
single-component Schmidt law (Schmidt 1959), with a dependence on gas
surface density  alone.  Plots of  the mean HII
region surface brightness in each annulus, which traces the mean star
formation rate per unit area in regions where star formation is
actually occuring,  show only a
modest variation with both radius and azimuthally--averaged gas surface
density (see Figure 2).  The abrupt steepenings observed in the
azimuthally--averaged profiles must therefore be almost entirely  due
to sharp declines in the {\it covering factor} of star formation in the
outer disk.  

\subsection{Does Gravitational Instability Drive Outer Disk Star Formation?}

The large--scale star formation process in galaxies is intimately
connected to the formation of gravitationally--bound gas complexes, and
giant molecular clouds (eg. Larson 1992).  The instability driving
cloud formation  may well be gravitational; in this case the Toomre-$Q$
criterion (Toomre 1964) yields a critical gas surface density above
which one expects local instability to axisymmetric perturbations.  For
an infinitely thin, one component isothermal gas disk, the critical gas
surface density above which self-gravity overcomes shear and pressure
is given by \begin{equation}
\Sigma_{crit}=\frac{{\alpha\sigma}{\kappa}}{\pi{G}} \end{equation}
where  $\alpha$ is a constant of order unity, $\sigma$ is the velocity
dispersion of the gas  and $\kappa$ is the epicyclic
frequency\footnote{The epicyclic frequency is defined as $\kappa=
\sqrt{2}~{\frac{V}{R}}\sqrt{1+{\frac{R}{V}}~{\frac{dV}{dR}}}$.}.
Following Kennicutt (1989), we adopt $\alpha=$0.67.

Published rotation curves and HI velocity dispersions exist for all
three galaxies, and can be used to calculate the radial variation of
$\Sigma_{crit}$.   Both NGC~628 and NGC~1058 have very  low inclinations
however, which means that the amplitude and shape of their rotation
curves are somewhat uncertain; for this reason, we will not consider them
further here.  For NGC~6946, we adopt the rotation curve of Sofue
(1996), normalised using the inclination derived by Carignan et al
(1990).   We made two estimates of the critical gas surface density,
one assuming a constant gas velocity dispersion (6~km~s$^{-1}$, for
consistency with Kennicutt 1989) and the other including the radial
variation determined directly from the HI observations (Kamphuis
1993).   Figure 3 shows  the radial variation of the ratio of the
azimuthally-averaged, deprojected  total gas surface density to the
critical gas surface density, calculated for each case.

The ratio of the observed gas density to the estimated critical density
is approximately constant across the range of radii probed
by our observations (the maximum variation is only a factor of 2 over
two optical radii).   Considering the likely uncertainties in the input
data which enter into the calculation -- eg.  rotation curves, velocity
dispersions, gas surface densities (specifically the possible variation
of the CO--H$_2$ conversion factor, and the existence of cold molecular
gas)  -- this result is even more compelling.

When a constant velocity dispersion is assumed, NGC~6946's
disk should become stable exactly at the optical radius.  The gas in
the extreme outer disk hovers very near the instability limit out to
two optical radii ($\Sigma_{gas}/\Sigma_{crit}$~$\sim$~0.8--1, or
1.5~$\lesssim$~Q~$\lesssim$~1.9).  The sharp decline in
azimuthally--averaged star formation rate is observed to begin at
0.8~R$_{25}$; if this behavior is to be ascribed to a transition in
the stability of the underlying gas disk, then either the value of
$\alpha$ or the velocity dispersion (or the combination) needs to be
increased by approximately 20\%.  On the other hand, when the observed
radial variation of the velocity dispersion is used in the calculation,
we find that the disk hovers just below the
instability limit at all radii
($\Sigma_{gas}$/$\Sigma_{crit}$~$\sim$~0.7--0.9, or
1.7~$\lesssim$~Q~$\lesssim$~2.1).  In this case, the inner and outer
disks are indistinguishable in terms of their local stability, and no
critical radius can be defined;  a Q-threshold  therefore
cannot provide an explanation for the steep declines in the covering
factor of star formation observed in the outermost parts of the disk.

Our analysis for NGC~6946 suggests that the extreme outer disk  
probably lies close enough to
the Q-stability limit so that processes such as swing amplification can
operate and perhaps  trigger star formation locally (eg.  Larson
1992).  While gravitational instability considerations may thus be able
account for the existence of star formation at large radii, only the
assumption of a constant gaseous velocity dispersion provides a  basis
for understanding the much reduced rates.  The sizes of the star
formation complexes and the spacing of the spiral arms may also be
difficult to reconcile with the gravitational instability model, since
the theoretically most unstable wavelength is much larger, and much
smaller, respectively, than those features which are observed in the extreme
outer disk (Ferguson et al 1998a).

\subsection{Star Formation in Flaring Gas Disks}

A possible alternative explanation for the observed behavior could be an
intrinsic correlation between azimuthally--averaged star formation rate
and gas {\it volume\/} density, combined with a vertical flaring of the
gas disk, such that the transformation between gas surface density and
volume density varies with galactocentric radius (see also Madore et al
1974). It is well known that gaseous disks  
exhibit an increase of
scaleheight at large radius (eg. Merrifield 1992, Olling 1996).  HII
regions, however, are confined to  a rather thin plane
with a constant scaleheight (eg.  Lockman, Pisano \& Howard 1996).  The
combination of an increase in gas scale height with radius with the slow
decline of gas surface density could conspire to produce a rapid
decline in areal star formation rate between the inner and the outer
disks.   The magnitude of the flarings required to  
reproduce the behavior of the large--scale star formation rate
do not appear unreasonable; a quantitative discussion is
given in Ferguson et al
(1998b).

\acknowledgements{We thank Jerry Sellwood,  Rob Kennicutt and Jean--Rene
Roy for interesting and useful discussions.  This
project was supported by an Amelia Earhart Fellowship from Zonta
International (AMNF), NASA grant NAGW-2892 (RFGW) and NASA contract NAS
7-1260 to JPL (JSG). }

\newpage

\begin{figure}
\figurenum{1a}
\caption{H$\alpha$+[NII] continuum--subtracted image of 
NGC~628; the optical radius
(R$_{25}$) is marked.}
\end{figure}
\begin{figure}
\figurenum{1b}
\caption{H$\alpha$+[NII] continuum--subtracted image of NGC~1058; the optical radius
(R$_{25}$) is marked.}
\end{figure}
\begin{figure}
\figurenum{1c}
\caption{H$\alpha$+[NII] continuum--subtracted image of NGC~6946; the optical radius
(R$_{25}$) is marked.}
\end{figure}

\begin{figure}
\figurenum{2}
\plotone{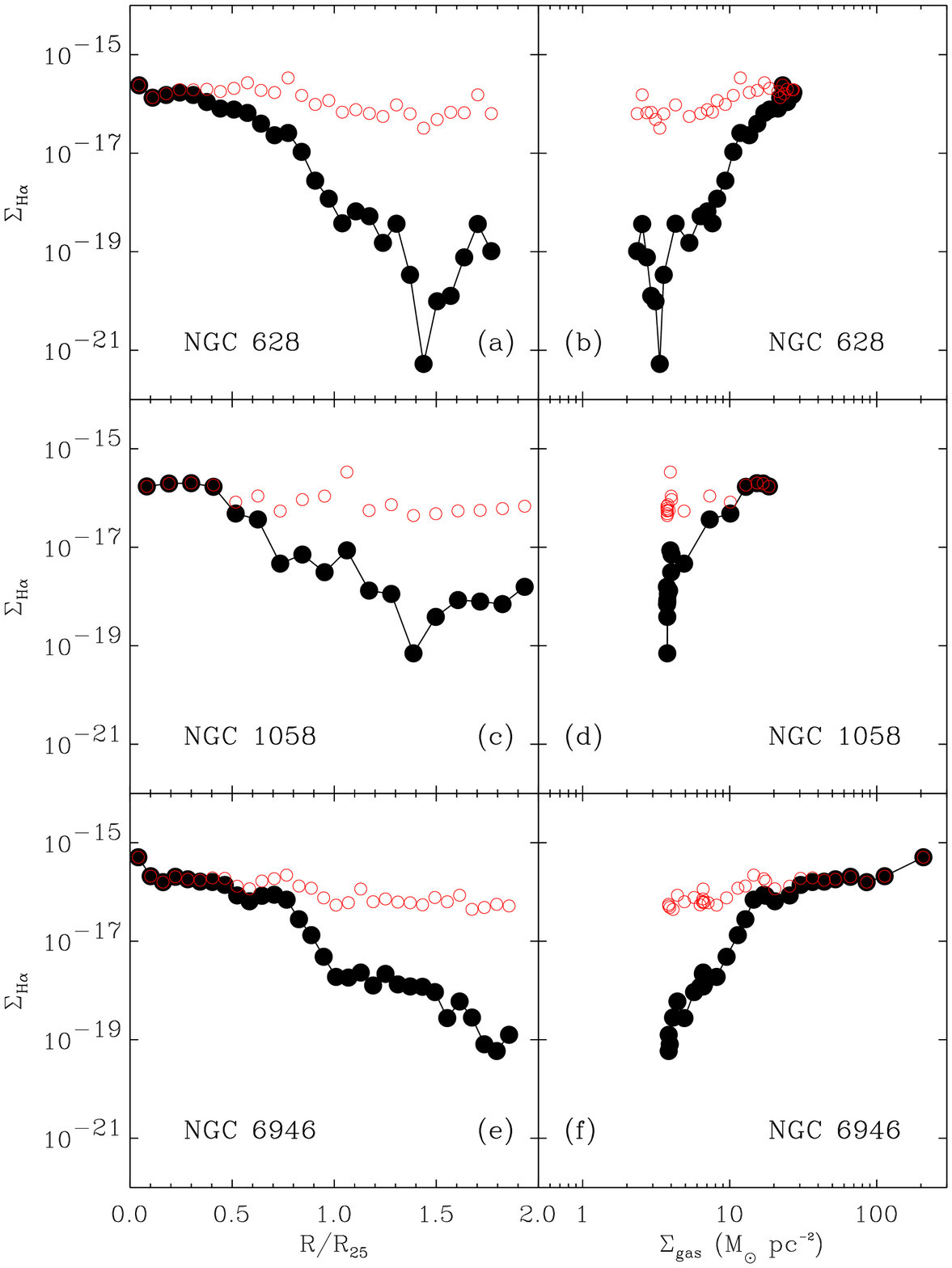}
\caption{(Left panel) The radial variation of the
deprojected, azimuthally-averaged  H$\alpha$ surface brightness,
$\Sigma_{H\alpha}$ (in erg~s$^{-1}$~cm$^{-2}~/\Box^{\arcsec}$) as
determined via elliptical aperture photometry (solid circles). (Right
panel) The variation of $\Sigma_{H\alpha}$ with the
azimuthally-averaged total gas surface density, $\Sigma_{gas}$, where
both quantities have been determined as a function of galactocentric
radius in a series of elliptical annuli (solid circles). Also plotted
are the mean HII region surface brightnesses, which trace the mean $`$local'  star formation rate per unit area (open circles).}
\end{figure}

\begin{figure}
\figurenum{3}
\plotone{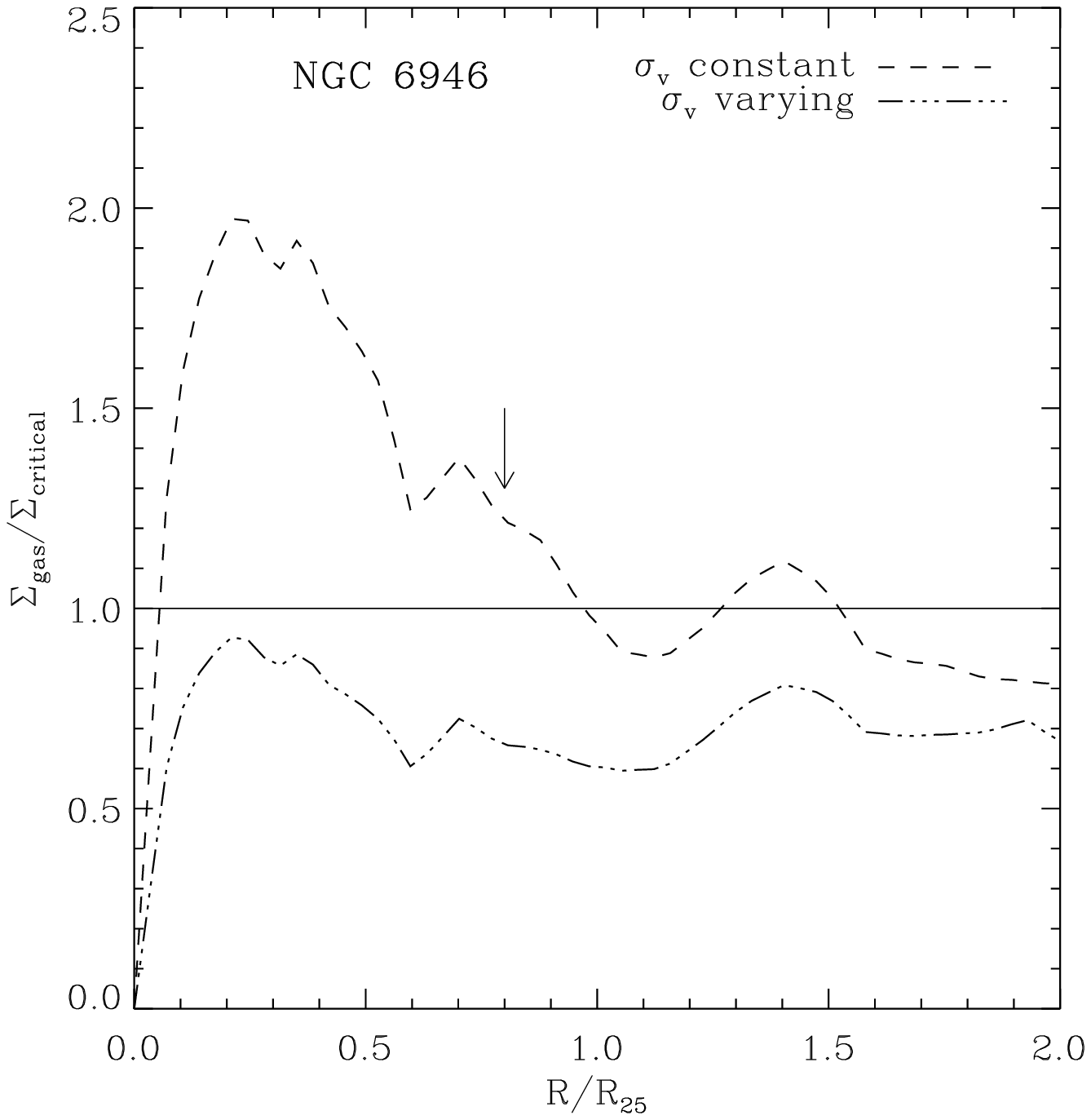}
\caption{The radial variation of the ratio of the
azimuthally--averaged (observed) gas surface density to the critical
surface density for gravitational instability, calculated using both a
constant velocity dispersion (dashed line) and a radially-varying one
(dashed-dotted line).  The solid horizontal line indicates the value of
the ratio above which instability is expected.  The small arrow indicates
the point in the disk where the azimuthally--averaged star formation
starts to decline sharply. }
\end{figure}

\end{document}